\documentclass[12pt,a4paper]{article}       

\usepackage{xcolor}
 \usepackage[skins,theorems]{tcolorbox}
\tcbset{highlight math style={enhanced,
  colframe=red,colback=white,arc=0pt,boxrule=1pt}}
  \usepackage[bookmarksopen, bookmarksnumbered, bookmarksopenlevel=2]{hyperref}
 \usepackage[UKenglish]{babel}
 \usepackage[toc,page]{appendix}
 \usepackage{amsmath}
 \usepackage{amssymb}
 \usepackage{graphicx}
 \usepackage{hhline}

 \usepackage[bf]{caption}
\usepackage{cite}
\usepackage[vcentermath]{youngtab}
\usepackage{geometry}
\usepackage{slashed}
\usepackage{stackrel}
\usepackage{mathtools}
\usepackage{cancel} 
\usepackage{multirow}
\usepackage[margin=0pt,font=small,labelfont=normalfont,skip=22pt]{subcaption}

\usepackage{empheq}
\usepackage{arydshln}

\newcommand{\bqa}{\begin{eqnarray}}
\newcommand{\eqa}{\end{eqnarray}}

\newenvironment{eqn*}{\begin{equation*}\begin{aligned}}{\end{aligned}\end{equation*}\noindent}
\hypersetup{
    pdftitle={},
    pdfauthor={},
    pdfsubject={}
}
\numberwithin{equation}{section}
\numberwithin{table}{section}

\makeatletter

\definecolor{dgreen}{rgb}{0,0.45,0.2}
\definecolor{dblue}{rgb}{0,0.0,0.5}

\newcommand{\be}{\begin{equation}}
\newcommand{\ee}{\end{equation}}
\newcommand{\beq}{\begin{equation}}
\newcommand{\eeq}{\end{equation}}
\newcommand{\ba}{\begin{aligned}}
\newcommand{\ea}{\end{aligned}}

\newcommand{\bea}{\begin{eqnarray}}
\newcommand{\eea}{\end{eqnarray}}

\newcommand\bi{\begin{itemize}}
\newcommand\ei{\end{itemize}}

\def\unit{{1\kern-.65ex {\rm l}}}
\def\1{{1\kern-.65ex {\rm l}}}

\newcount\hour \newcount\minute
\hour=\time \divide \hour by 60
\minute=\time
\def\now{%
\ifnum \hour<13
  \ifnum \hour=0 \advance \hour by 12 \number\hour:\else \number\hour:\fi%
     \ifnum \minute<10 0\fi%
     \number\minute%
\ A.M.%
\else \advance \hour by -12 \number\hour:%
  \ifnum \minute<10 0\fi%
  \number\minute%
  \ P.M.%
\fi%
}

\makeatother

\begin{document}

\begin{titlepage}
\begin{center}
\rightline{\small }

\vskip 15 mm

{\large \bf
Ray-Singer Torsion, Topological Strings and Black Holes} 
\vskip 11 mm

 Cumrun Vafa

\vskip 11 mm

{\it Jefferson Physical Laboratory, Harvard University, Cambridge, MA 02138, USA}

\end{center}
\vskip 17mm

\begin{abstract}
Genus one amplitude for topological strings on Calabi-Yau 3-folds can be computed using mirror symmetry:  The partition function at genus one gets mapped to a holomorphic version of Ray-Singer torsion on the mirror Calabi-Yau.  On the other hand it can be shown by a physical argument that this gives a curvature squared correction term to the gravitational action.  This in paticular leads to an effective quantum gravity cutoff known as the species scale, which varies over moduli space of Calabi-Yau manifolds.  This resolves some of the puzzles associated to the entropy of small black holes when there are a large number of light species of particles.  Thus Ray-Singer torsion, via its connection to topological strings at genus one, provides a measure of light degrees of freedom of four dimensional ${\cal N}=2$ supergravity theories.

{\it Based on a talk given on May 12th, 2023 at the Singer Memorial Conference, MIT.}
\end{abstract}

\vfill
\end{titlepage}

\section*{Isadore Singer}
Isadore Singer was not only a great mathematician, but also a science visionary who appreciated the importance of deep links between modern mathematics and theoretical physics.  Moreover, he played an active role in fostering the development of a new branch of science which is at the interface of both fields.  On the one hand this  equips physicists with powerful mathematical tools to explore laws of nature at its most fundamental level and on the other hand it provides physical insights to mathematicians in uncovering and connecting different areas of mathematics.

I have been fortunate to have also known Is as a friend and to discuss these topics with him frequently.  Moreover I have many fond memories of our joint efforts which helped establish the Simons Center for Geometry and Physics in Stony Brook.  

The talk I presented in the Singer Memorial conference at MIT, was motivated by giving a concrete example of showing this deep back and forth connection between physics and mathematics.   The example I presented uses a beautiful work of Singer (Ray-Singer torsion \cite{ray1971rtorsion,ray1973analytic}) and its relation to Gromov-Witten invariants (topological strings) via mirror symmetry.  All these ideas have found applications in the past year in the context of black hole physics and positive potentials that can appear in physical theories of quantum gravity \cite{vandeHeisteeg:2022btw,vandeHeisteeg:2023uxj}, which I review in this paper.

\section{Introduction}
Ray-Singer Torsion \cite{ray1971rtorsion,ray1973analytic}, a concept intertwining mathematics and physics, has recently become important in understanding quantum gravity and string theory. We first briefly review
its historical development and then discuss its significance in modern theoretical physics through a particular example.

\section{Ray-Singer Torsion in Compact Kahler and Calabi-Yau Manifolds}

 Among the many important works that Singer has done, here I will describe an application of an object which arises in physics and closely parallels what Ray and Singer discovered.  They considered \cite{ray1971rtorsion,ray1973analytic} an arbitrary Riemannian manifold M (together with a vector bundle E), and defined

$$T(M,E)=\prod_p\rm det(\Delta_p)^{\frac{-(-1)^p p}{2}}$$
where $\Delta_p$ is the Laplacian acting on $p$-forms, and the determinant is made precise by zeta function regularization.
It was conjectured by Ray and Singer that this invariant is an analytic version and equivalent to the Reidemeister torsion \cite{reidemeister1935homotopieringe}.
This was later proven by Cheeger \cite{cheegerAnalyticReidemeister,cheeger1979analytic} and Muller \cite{muller1978analytic}.

In this talk I will discuss a closely related object, which is a holomorphic version of Ray-Singer torsion that arises for compact Kahler manifolds (and more specifically Calabi-Yau manifolds).  This is ``invariant'', in the sense that it does not depend on the Kahler metric, but does depend on the complex structure.

We will encounter a particular combination of the holomorphic Ray-Singer torsion:

$$ \prod_{p,q=0}^n({\rm det} \ \Delta_{p,q})^\frac{(-1)^{p+q}pq}{2}$$
where $\Delta_{p,q}$ denotes Laplacian acting on $(p,q)$ forms.  We will explain how this ends up `counting’ the number of light particles by invoking the physics of black holes.

My main emphasis in this note is to provide an example of how different ideas in physics and mathematics intermingle to lead to insights for both fields.
On the physics side we explore aspects of
black holes and their information content, and on the joint mathematical and physical side consider Gromov-Witten invariants (also known as topological string amplitudes) on Calabi-Yau threefolds.  Also mirror symmetry between Calabi-Yau threefolds enters this story, where the Gromov-Witten invariants on a given Calabi-Yau are computed by going to the mirror and doing the corresponding computation there, which turns out to be related to Ray-Singer torsion.
Thus the holomorphic version of Ray-Singer torsion ends up computing Gromov-Witten invariants for genus 1 curves.  On the physics side, insights on the structure of light fields in quantum gravitational theories, including bounds on the nature of positive potentials $V>0$ that can appear in physical theories, all get connected to Ray-Singer torsion.

This talk is based on old and new works.  In particular the old work with Bershadsky, Cecotti and Ooguri \cite{bershadsky1994kodaira} and the new work with
van de Heisteeg, Wiesner and Wu \cite{vandeHeisteeg:2022btw,vandeHeisteeg:2023ubh}.
Road-map for the rest of this note is as follows.  In section 3 we explain the sense in which the Planck scale is viewed as the short distance cutoff of quantum gravity theories.  In section 4 we briefly explain the Bekenstein-Hawking entropy of black holes.  In section 5 we explain why having many light species raises a puzzle and a resolution is offered by replacing the Planck scale cutoff with the Species scale cutoff.  In section 6 some aspects of string dualities are reviewed with particular attention to how the light states appear as we approach boundaries of moduli spaces.  In section 7 we review the relevance of one-loop topological strings (genus 1 Gromov-Witten invariants) to certain $R^2$ curvature corrections in the action. Moreover it is shown how mirror symmetry can be used to relate this to a holomoprhic version of Ray-Singer torsion and how this leads to its computation.  We use this to find the dependence of the species scale on  moduli which leads to interesting bounds on the allowed potential in physical theories.  We also explain why bounds on the shape of potential is an important topic for theoretical physics today.  In section 8 we end with some concluding remarks.

\section{Planck Scale as a UV Quantum Gravity Cutoff}

 In quantum gravity, the leading term of the action is the Einstein-Hilbert action.  However we also expect corrections to the Einstein-Hilbert action:
$$ S = \int d^dx \, \sqrt{-g} \left[\frac{M_{pl}^{d-2}}{2} \left(R +\sum_n  \,\frac{R^n}{M_{pl}^{2n-2}}+\dots  \right) \right]$$
Here we have put the fundamental mass scale, the Planck mass $M_{pl}$ to make the dimensions right, and 
we expect the coefficients  of each term to be of $O(1)$, which we have not explicitly written.
Note that curvature has dimension of mass squared.  Of course not all higher derivative terms will have coefficients of $O(1)$, but that should be the generic case.
This implies that when curvature $R\sim M_{pl}^2 $  the perturbation series breaks down and the geometric description given by the leading Einstein-Hilbert term, breaks down.

We should thus view
$\Lambda_{UV} \sim M_{pl}$
as the high energy ultra-violet cutoff where geometric description breaks down.  This means that since length has inverse dimension to mass, the Planck length
$l_{pl} = M_{pl}^{-1} $
should be viewed as the shortest geometric length scale in the theory.

\section{Basics of Black Hole Entropy}
Black hole is one of the most enigmatic objects not only in nature, but also for theoretical physics.  One of the most important properties of black holes, which was discovered half a century ago, was the work of
Bekenstein and Hawking who found that when you include quantum corrections a black hole can be consistent only if it represents not one but many potential objects; i.e. that it has entropy. Moreover the entropy of the black hole is proportional to the area of its event horizon (which is the point beyond which no light can classically escape to infinity). This 
was originally motivated by the classical observation of the fact that the area of the black hole horizons increase over time and made precise by combining with semi-classical properties of black holes.  More precisely if we denote by $A$ the area of the event horizon measured in Planck units, the entropy of the black hole is given by
$$S_{BH}=\frac{1}{4}A\sim \bigg(\frac{R_{BH}}{l_{pl}}\bigg)^{d-2}$$
The entropy signifies the number of micro-states of a black hole.  If we assume a black hole is made of $N$ independent constituents (so that the total number of states goes as $e^N$) then this is the same as expecting $N$ to grow with the radius of the black hole by
$$
S_{BH}\sim N\sim R_{BH}^{d-2}.$$

We will now see why this leads to a contradiction with the expectation that the Planck scale is the quantum gravity cutoff, when we have many light species of particles.

\section{Light Species Puzzle for Black Holes}

Suppose we have $N\gg 1$ of light ($m\ll M_{pl} $) states.   Then for any black hole bigger than Planck scale (so that we can ignore the corrections to the Einstein-Hilbert action), we would expect the entropy of black hole $S_{BH}$ satisfies
$$S_{BH}>N,$$
as the light particles can constitute the degrees of freedom of the black hole.

On the other hand, if we assume the size of the smallest black hole describable reliably by the Einstein-Hilbert action is Planck length, as one would naively expect based on quantum gravity, this would suggest $S\sim 1$ for this black hole which is inconsistent with the above entropy bound.
This is the species problem which we now review.
Species scale \cite{Dvali:2007hz} was offered as a solution to this problem.

\subsection{                                   Species Scale}           
Suppose we have a large number $N$ of light species. 
Then it better be that gravitational action breaks down before we get to Planck scale:  
$$S_{BH}\sim 1 \quad {\rm for} \ R_{BH}\sim l_{Pl}$$
because this would not accommodate $N$ degrees of freedom.  Indeed the smallest size black hole compatible with such an entropy has a larger radius (using Bekenstein-Hawking entropy):
$$\bigg(\frac{R_{BH}^{min}}{l_{pl}}\bigg)^{d-2}\sim N\Rightarrow \frac{R_{BH}^{min}}{l_{pl}}\sim N^{\frac{1}{d-2}}.$$
How could this be explained?  A natural explanation of this would be that
higher derivative corrections to the action are important and compete at the mass scale 
$$\Lambda_s=\frac{1}{R_{BH}^{min}}\sim \frac{M_{pl}}{N^{\frac{1}{d-2}}} .$$
$\Lambda_s$ is called the species scale.
We thus have a new UV cutoff for gravity:       $\Lambda_s< M_{pl}$.

\begin{figure}[ht]
    \centering
    \includegraphics[width=.5\textwidth]{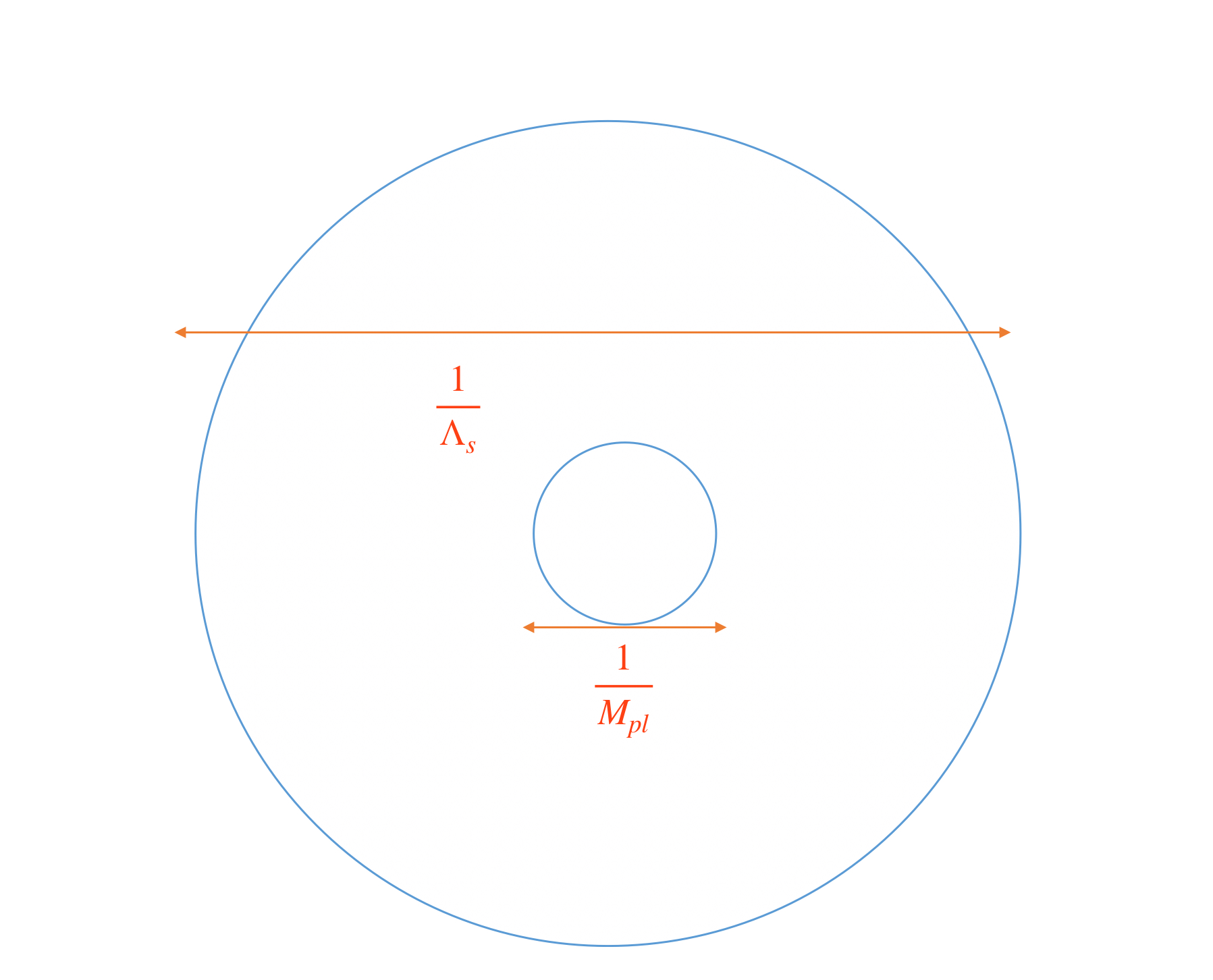}
    \caption{The minimal size of the black hole which can be reliably treated using the effective action is $1/\Lambda_s$ which is larger than the Planck length $1/M_{pl}$.}
    \label{fig:enter-label}
\end{figure}

In other words
$$\bigg(\frac{M_{pl}}{ \Lambda_s}\bigg)^{d-2}= N$$

 is a measure of the number of light degrees of freedom in the theory.   The smaller the $\Lambda_s$
the more the number of light states.  We thus expect $\Lambda_s$ to replace $M_{pl}$ in higher derivative correction terms:

$$ S = \int d^dx \, \sqrt{-g} \left[\frac{M_{\rm pl}^{d-2}}{2} \left(R +\sum_n  \,\frac{R^n}{\Lambda_s^{2n-2}}+\dots  \right) \right]$$

\section{Dualities and Moduli Space}                            

One of the basic features of string theory is the existence of duality symmetries in the theory.  This means that if you change the parameters in the theory to extreme values, you get a new description of the theory.  The mechanism for this is that at extreme values of parameters light tower of particles appear that lead to a new weakly interacting description of the theory.
The parameters of the theory are typically moduli of manifolds where strings are propagating in.  Let us  parameterize the moduli space of compactified space by the field $\phi$ (which can be viewed as a coordinate system on the moduli space).  Moreover there is a natural metric $g_{ij}(\phi)$ on this space arising from the kinetic term in the action:
$$S=\frac{1}{2} \int g_{ij}(\phi) \ \nabla \phi^i \cdot \nabla \phi^j+...$$

\begin{figure}[ht]
    \centering
    \includegraphics[width=\textwidth]{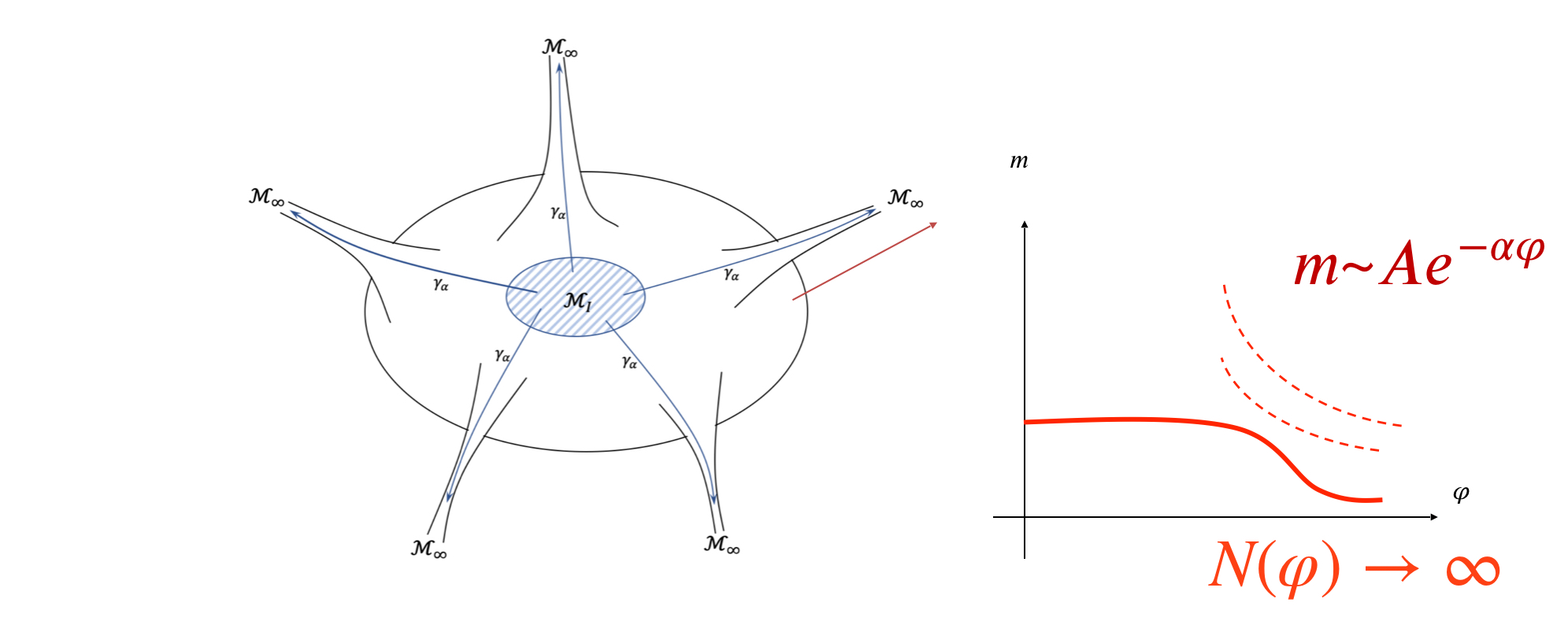}
    \caption{At extreme corners  of the moduli space a light tower of particle appears whose mass scales down exponentially with distance.  The effective number of light fields $N$ increases as we approach the boundaries.  These light fields lead to the dual description of the theory.}
    \label{fig:enter-label}
\end{figure}

 Example:  Consider compactification of IIA string on Calabi-Yau 3-fold.  The Kahler and complex moduli of CY correspond to massless fields of the 4 dimensional theory.  Let us focus on Kahler moduli $k=e^{\varphi} k_0$.
 As  $\varphi \gg 1 \Rightarrow m\sim {\rm exp}\big[-\sqrt{\frac{3}{2}}\varphi \big]$
 these come from low lying eigenvalues of Laplacian (KK modes) in large volume.
Another limit is of the following type:  Consider K3 fibered CY3-fold, and consider the limit the volume of K3 fiber goes to zero.  In this limit we get tensionless strings
coming from 5-branes wrapped on shrinking K3:
$$Vol(K3)=e^{-\varphi}, \qquad \varphi \gg 1\Rightarrow m\sim {\rm exp}\big[ -\frac{\varphi}{ {\sqrt 2}}\big ]$$
where the tension of the string (in lower dimensional Planck units) is $T\sim m^2$.
In general the exponent of the decay of the mass with field space distance satisfies the bound:
$$
\sqrt{\frac{D-2}{(D-d)(d-2)}} \geq \alpha \geq \frac{1}{\sqrt{ (d-2)}} $$
where we consider theories compactified on a $D-d$ dimensional manifold down to $d$ dimensions.
In this bound $D\leq 11$ which is the maximum dimension for M-theory.  The lower range of the bound gets saturated when we get a tensionless string.

This implies the species scale will depend on the value of massless fields $\varphi$, i.e., $\Lambda_s(\varphi)$  (since as we change $\varphi$ the masses change and so the species scale could change).  We thus expect
$$ S = \int d^dx \, \sqrt{-g} \left[\frac{M_{\rm pl}^{d-2}}{2} \left(R +\sum_n  \,\frac{R^n}{\Lambda_s^{2n-2}(\varphi)}+\dots  \right) + \frac{1}{2}|\partial \varphi|^2+\dots\, \right]$$
We are interested in finding the dependence of $\Lambda_s$ on $\varphi$.
Let us focus on $d=4$ and the $R^2$ term, which is accompanied by $\frac{1}{ \Lambda_s^2(\varphi)}\sim N(\varphi)$. 
How to compute this term $\int d^4x \ N(\varphi)\cdot R^2$ ?

One can compute this term in string perturbation theory but then it will not be robust:  As we go to strong coupling regime of string theory, the corrections would be difficult to compute as we would need to include not only the contributions coming from all genera, but also non-perturbative contributions.
However, there is a magical $R^2$-type correction which {\it can} be computed using string perturbation techniques and that it receives no quantum corrections and it is exact!
To explain this we need to turn to topological strings.

\section{Topological Strings on CY 3-folds and dependence of Species scale on moduli}

Topological strings on Calabi-Yau 3-folds calculates the Gromov-Witten invariants, which roughly speaking `counts' holomorphic maps to the CY (for more precise definition see e.g. \cite{mirror_symmetry_ams}).  More specifically, the generating function of this invariant is the partition function of the topological string on the corresponding CY.

\begin{figure}[ht]
    \centering
    \includegraphics[width=\textwidth]{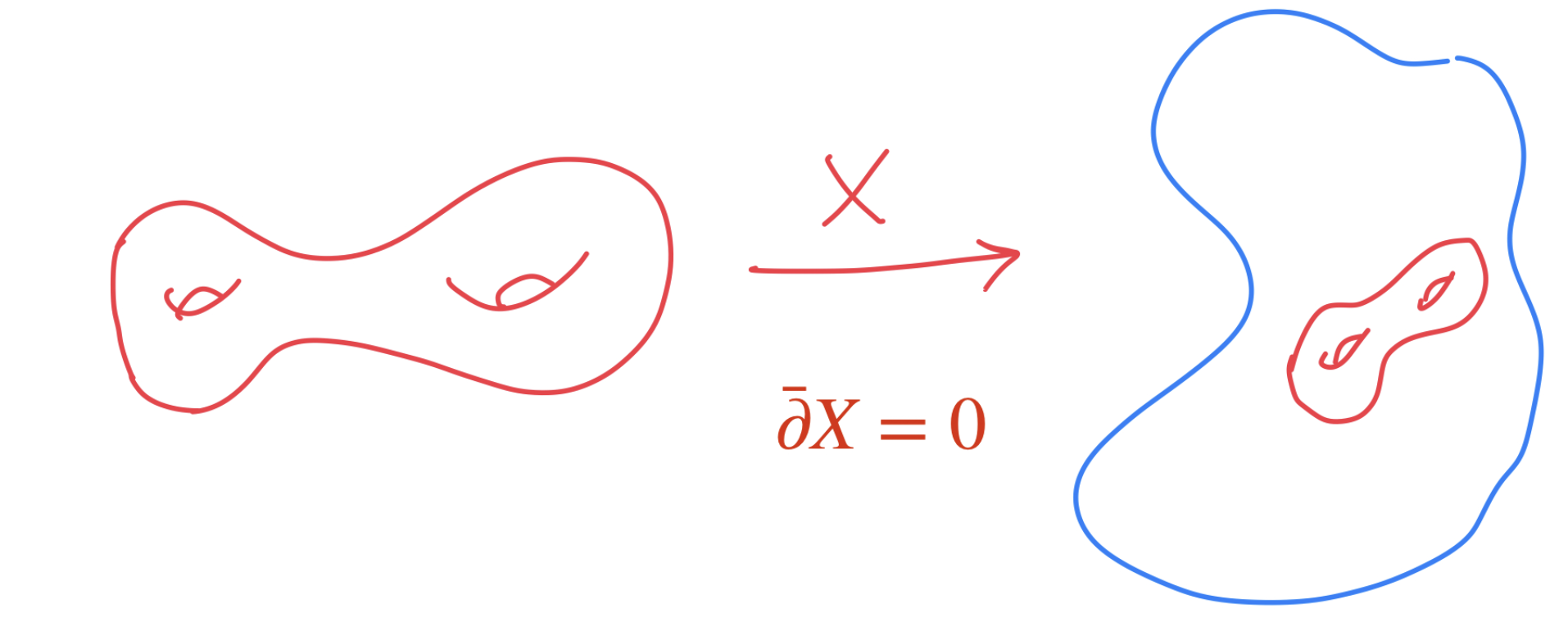}
    \caption{Gromov-Witten invariants `count' holomorphic maps from curves to Calabi-Yau manifolds.}
    \label{fig:enter-label}
\end{figure}
The topological string partition function at genus g is given by

$$F_g(k)=\sum_{\bar \partial X=0}{\rm exp}(-\int_{\Sigma^g}X^*(k))$$

Where  $k={\rm Kahler\  form}$ of CY and $\Sigma^g $ is the $ {\rm genus\ g\ curve}$.
The topological string partition function compute physical terms for superstring action.  In particular $F_1$ computes in the four dimensional action, the term given by
$$\int \ F_1(k)\cdot( R^{-})^2$$
where $R^{-}$ is the anti-self-dual part of the curvature.  Moreover this term is exactly computable as it receives no further corrections! In other words the genus 1 computation of the topological string yields the exact answer for this quantity! We can thus interpret this quantity in the four dimensional theory, since $\frac{1}{\Lambda_s^2}\sim N$, as implying:
$$F_1(k)\sim N(k).$$
So to complete the circle of ideas we need to know how to compute $F_1$, the genus one topological string amplitude?
Here we can use mirror symmetry between Type IIA strings on $CY3$ with Type IIB on Mirror $CY3’$ (See \cite{mirror_symmetry_ams}
for more detail on this).  In particular note that $$h^{p,q}(CY)=h^{p,3-q}(CY').$$
This maps the Kahler parameters of $CY$ $k$ (parameterized by $h^{1,1}$) to complex moduli $t$ of $CY'$ (parameterized by $h^{1,2}$).  Therefore
Topological String on $CY$ which is parameterized by Kahler deformations gets mapped to complex deformation of $CY'$.
$$F_g(k)\Rightarrow F_g(t)$$
The latter is computed using the 
topological B-model, which is the `quantum Kodaira-Spencer theory of gravity’ \cite{bershadsky1994kodaira}.  This in turn is much simpler than counting curves and is given in terms of a field theory computation, whose classical equations lead to Kodaira-Spencer equations for the deformation of complex structure.
The genus one amplitude for topological string thus gets mapped to one-loop corrections of the field theory, which in turn, as in all field theories, is given by certain combination of determinants.
$$F_1\ \ \ \  \ \ \ \ \ \Rightarrow  \ {\rm one\ loop\ correction}$$

\begin{figure}[ht]
    \centering
    \includegraphics[width=\textwidth]{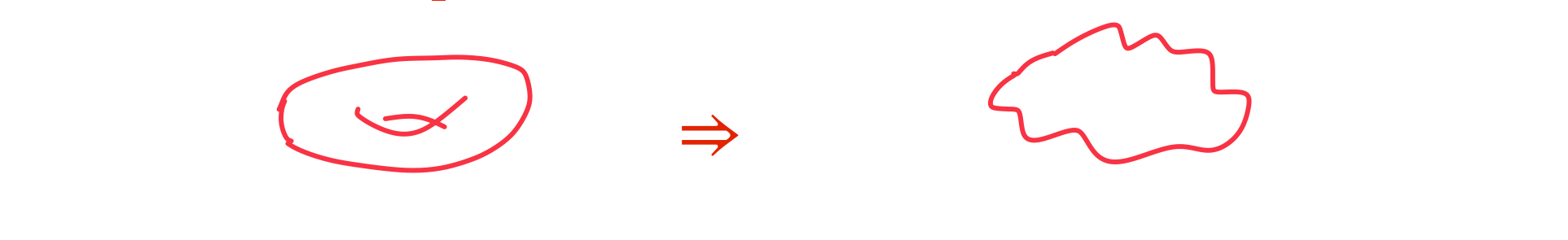}
    \caption{The genus 1 Gromov-Witten invariants get mapped using mirror symmetry to one loop amplitudes of Kodaira-Spencer theory, which in turn is given by a holomorphic version of Ray-Singer torsion.}
    \label{fig:enter-label}
\end{figure}
 This combination of determinants happens to be a combination of holomorphic Ray-Singer torsion:

$$ e^{F_1}=\prod_{p,q=0}^3({\rm det} \ \Delta_{p,q})^{\frac{(-1)^{p+q}pq}{2}}$$
which satisfies \cite{Bershadsky:1993ta}
$${\partial \bar \partial} F_1= \text{sTr}\  C \bar{C} -\frac{1}{12} G \cdot \chi \,, $$
where $C$ is a certain natural action of the tangent of moduli on the harmonic forms of the CY.  This result can also be seen \cite{bismut1986analysis1,bismut1986analysis2} as an application of Quillen’s work for curvature of the determinant line bundle).
Integrating the above equation yields
$$F_1 = \frac{1}{2}\left(3+h^{1,1}-\frac{\chi}{12}\right) K\,-\frac{1}{2}\log \det G +\log|f|^2\,,$$
Where K is the Kahler function on complex structure moduli space, $\chi$ is the Euler character of CY, G is the metric on moduli space and $f$  is a holomorphic function of moduli.

To fix everything we need to find the holomorphic function $f$ which is fixed by the behavior expected at the boundary (which in turn is determined by using mirror symmetry).

A specially easy case is the CY 3-fold given by a $Z_2$ quotient of the product of $K3$ and the elliptic curve where the $Z_2$ acts as inverstion on $T^2$ and acts freely on $K3$ taking the holomorphic 2-form to its negative.  This yields for
$CY^3=K3\times T^2/Z_2$:

$$F_1=-12 \ {\rm log} \big[\sqrt \tau_2 \eta(\tau)\eta(\bar \tau )\big]\sim N\sim \frac{1}{ \Lambda_s^2}$$

$$N\sim \tau_2\sim {\rm exp}(\sqrt 2\ \varphi) \ \ {\rm for}\ \tau_2 \gg 1 $$
consistent with what one expects for the behaviour of $\Lambda_s$ and $N$ near the boundary of moduli space.

From this example and many others we find the general picture that the species scale decays exponential at the boundaries and reaches a maximum in the interior of moduli space. Moreover, we find that the exponential fall off of $\Lambda_s$ with distance on moduli gives the expected exponent:

\begin{figure}[ht]
    \centering
    \includegraphics[width=\textwidth]{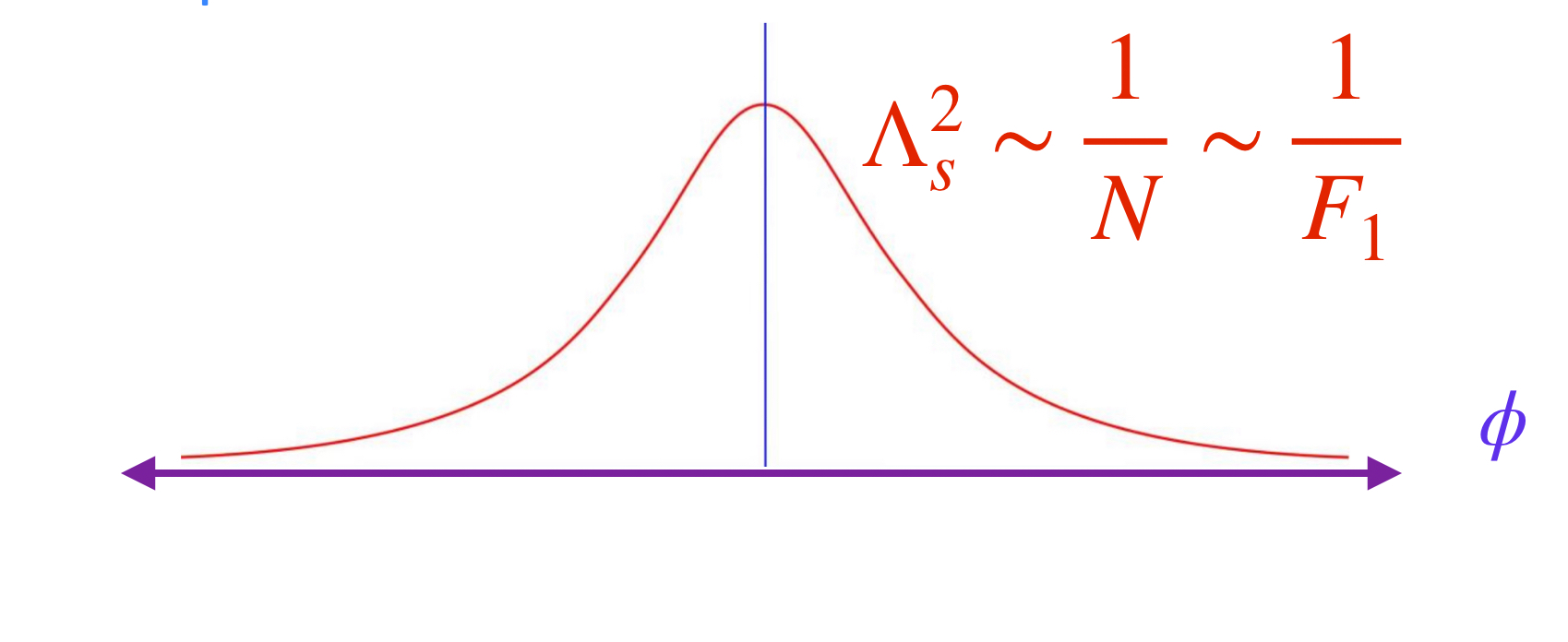}
    \caption{A typical behaviour of $\Lambda_s^2$ as a function of moduli fields:  exponential decay at infinity and a maximum of order 1 in the interior.}
    \label{fig:enter-label}
\end{figure}

\subsection{Application to Bounds on positive potentials}

One of the most challenging and phenomenologically interesting questions in quantum gravity is whether we can say anything about the nature of positive potentials $V>0$.  This is particularly difficult because such situations necessarily imply that supersymmetry is broken and we thus have no exact techniques to compute physical amplitudes.  However, for at least two reasons putting bounds on $V>0$ is crucial for cosmology:
An important question for early cosmology in the context of inflation is whether long ranged flat positive potentials $V>0$ can exist and if there are any restrictions on it.                              

\begin{figure}[ht]
    \centering
    \includegraphics[width=\textwidth]{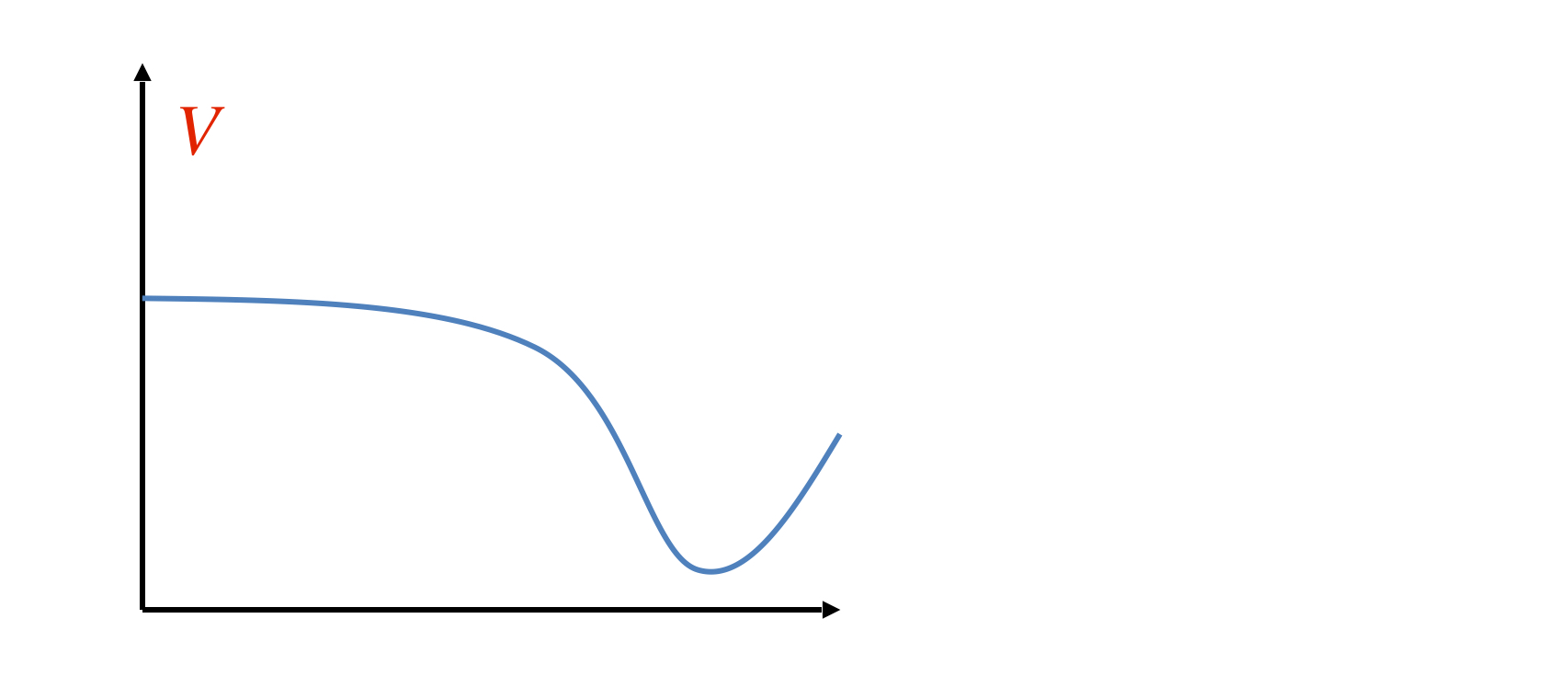}
    \caption{Long flat regions of potential are needed for generic inflationary model.}
    \label{fig:enter-label}
\end{figure}

We now argue that for flat regions of the potential
$$V(\varphi) < \Lambda^2_s(\varphi)$$
This can be argued as follows:
For flat regions of potential, the space is approximately de Sitter (homogenous space with positive constant curvature) and the maximal size of the space is given by the horizon size
$$R_{max}\sim \frac{1}{\sqrt{V}}$$
\begin{figure}[ht]
    \centering
    \includegraphics[width=\textwidth]{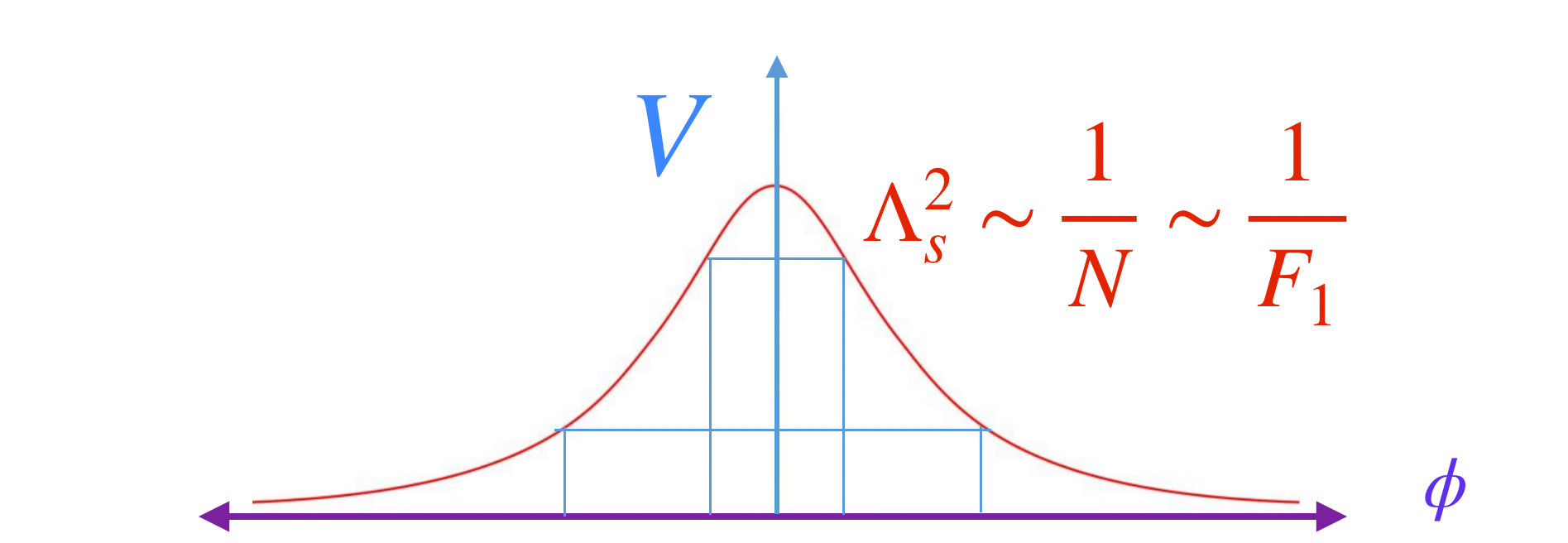}
    \caption{Using the exponential behaviour of the species scale near the boundaries we can get a bound on the width of regions with flat potential.}
    \label{fig:enter-label}
\end{figure}

As we already argued the minimum distance for which we can use the effective action is the species scale distance.  This leads to a bound on $V$:
$$R_{min}\sim \frac{1}{\Lambda_s}<R_{max}\sim \frac{1}{\sqrt{V}}\Rightarrow V<\Lambda_s^2$$

Using this, and the asymptotic nature of the species scale, we can get bounds on length of regions of flat regions for V from $V<\Lambda_s^2$.  In particular we learn that
high flat regions of potential lead to small range and the only way to get larger ranges is to decrease the potential:                         $$\Delta \varphi <a .{\rm log}\bigg[\frac{1}{V}\bigg]-b$$
where    
$$a\leq \sqrt{(d-2)(d-1)};  \ \  b\sim O(1)$$

\section{Conclusion}
We have seen that the species scale is an important concept in quantum gravitational theories and can depend on moduli of light fields. Topological strings and mirror symmetry lead to a computation of the species scale: the one-loop topological amplitude counts the light degrees of freedom which in turn is given by a holomorphic version of Ray-Singer Torsion.
This also leads to bounds on the range of flat regions of potential which scale logarithmically with the inverse height of the potential
\cite{vandeHeisteeg:2023uxj}.

The intertwined mathematical and physical ideas that were discussed here provide an example of how the old work of Singer and his vision continue to impact the development of physics and mathematics today.

\subsubsection*{Acknowledgments} 
I thank my collaborators, whose work I presented here.
This work is supported by a grant from the Simons Foundation (602883,CV), the DellaPietra Foundation, and by the NSF grant PHY-2013858.

\bibliographystyle{jhep}
\bibliography{sample}

\end{document}